\documentclass[fleqn,10pt]{wlscirep}
\usepackage[utf8]{inputenc}
\usepackage[T1]{fontenc}
\usepackage{bm}
\usepackage{multirow}

\usepackage{xcolor}

\def\change#1{#1}

\title{Ising machines: Hardware solvers for combinatorial optimization problems}

\author[1,2,3]{Naeimeh~Mohseni}
\author[4,*]{Peter~L.~McMahon}
\author[5,1,6,7,8,$\dagger$]{Tim~Byrnes}

\affil[1]{State Key Laboratory of Precision Spectroscopy, School of Physical and Material Sciences, East China Normal University, Shanghai 200062, China}
\affil[2]{Max-Planck-Institut f{\"u}r die Physik des Lichts, Staudtstrasse 2, 91058 Erlangen, Germany}
\affil[3]{Department of Physics, University of Erlangen-Nuremberg, Staudtstr. 5, 91058 Erlangen, Germany}
\affil[4]{School of Applied and Engineering Physics, Cornell University, Ithaca, NY 14853, USA}
\affil[5]{New York University Shanghai, 1555 Century Ave, Pudong, Shanghai 200122, China}
\affil[6]{NYU-ECNU Institute of Physics at NYU Shanghai, Shanghai 200062, China}
\affil[7]{National Institute of Information and Communications Technology, Tokyo 184-8795, Japan}
\affil[8]{Department of Physics, New York University, New York, NY 10003, USA}
\affil[*]{pmcmahon@cornell.edu}
\affil[$\dagger$]{tim.byrnes@nyu.edu}

\begin{abstract}
Ising machines are hardware solvers which aim to find the absolute or approximate ground states of the Ising model. The Ising model is of fundamental computational interest because it is possible to formulate any problem \change{in the complexity class NP} as an Ising problem with only polynomial overhead. A scalable Ising machine that outperforms existing standard digital computers could have a huge impact for practical applications for a wide variety of optimization problems.  In this review, we survey the current status of various approaches to constructing Ising machines and explain their underlying operational principles. The types of Ising machines considered here include classical thermal annealers based on technologies such as spintronics, optics, memristors, and digital hardware accelerators; dynamical-systems solvers implemented with optics and electronics; and superconducting-circuit quantum annealers. We compare and contrast their performance using standard metrics such as the ground-state success probability and time-to-solution, give their scaling relations with problem size, and discuss their strengths and weaknesses.
\end{abstract}

\begin{document}

\flushbottom
\maketitle

\thispagestyle{empty}

\section*{Key points}

\begin{itemize}
    \item Dedicated hardware solvers for the Ising model are of great interest due to the many potential practical applications and the end of Moore's law which motivate alternative computational approaches. 
    
    \item Three main computing methods that Ising machines employ are classical annealing,  quantum annealing, and dynamical system evolution. A single machine can operate based on multiple computing approaches.  

    \item  \change{Today, Ising hardware based on classical digital technologies are the best performing for common benchmark problems.   However, the performance is problem dependent and alternative methods can perform well for particular classes of problems.} 
    \item  For particular crafted problem instances, \change{quantum approaches have been observed to have a superior performance over classical algorithms,} motivating quantum hardware approaches and quantum-inspired classical algorithms.  
    \item Hybrid quantum-classical and digital-analog algorithms are promising for future development; they may harness the complementary advantages of both. 
   
\end{itemize}

\section*{Introduction}
Computers have had an enormous impact on society, infiltrating almost every aspect of our daily lives.  One type of problem that conventional computers have particular difficulty in solving are hard combinatorial optimization problems. Such problems typically involve finding an optimal configuration, defined by a cost function, among a very large number of potential candidate configurations. Examples of such problems include the travelling salesman problem, Boolean satisfiability problems (i.e. SAT problems), MaxCut, to name a few.  In a practical setting, such combinatorial optimization problems are of relevance to a wide variety of applications such as planning, logistics, manufacturing, financial portfolio management, computer vision, artificial intelligence, machine learning, bioinformatics, drug design, and a variety of chemical and physical materials problems \cite{lucas2014ising, tanahashi2019application, smelyanskiy2012near,hauke2020perspectives}.  In many cases, such combinatorial optimization problems are instances of NP-complete problems, which represent the hardest problems within the NP class. A well-known result states that it is possible to map any problem in NP to a NP-complete problem in polynomial time \cite{karp1972reducibility,mezard1987spin}. 
In this sense, if there were a way of solving any combinatorial optimization problem in the NP-complete class with an improvement over conventional computing methods,
then this would have an enormous impact to a large number of practical applications. 

\begin{figure}[t]
\begin{center}
\includegraphics[width=\linewidth]{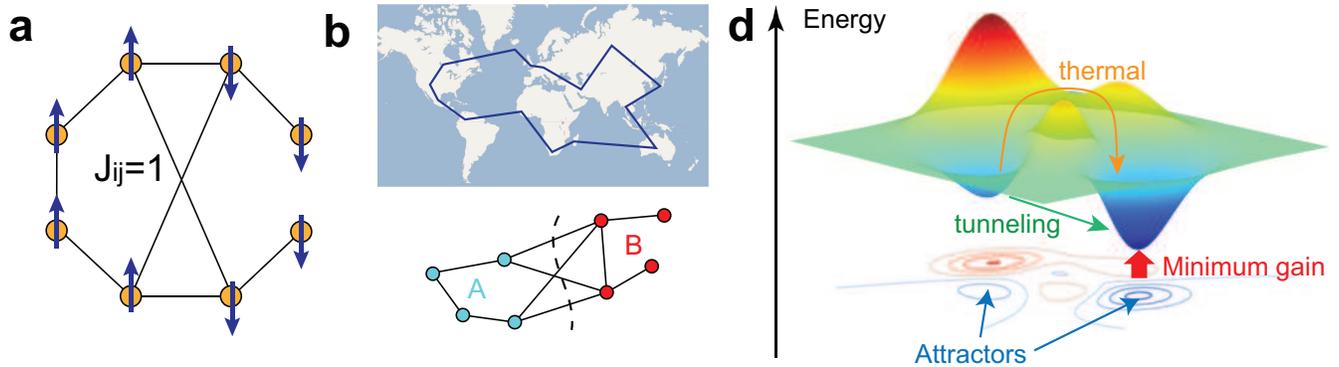} 
\end{center}
\caption{{\bf The Ising model, combinatorial problems, and its energy landscape.}  (a) An example of an 8 spin Ising model.  On each node is a two valued spin (arrows). Edges correspond to assigning an energy $ J_{ij} = 1$. (b) Examples of combinatorial optimization problems:  the travelling salesman problem (top) and MaxCut (bottom).  In the travelling salesman problem, the aim is to find the shortest possible route that visits each city exactly once and returns to the origin city. For MaxCut, the graph is 
equivalent to the Ising model in (a) and the the optimal division is indicated by the dashed line.  The travelling salesman problem can be mapped onto the Ising model by encoding the information of the city and its route ordering as a spin variable.  The total number of spins required is the square of the number of cities.  (c) Schematic energy landscape of the Ising model and various mechanisms used in Ising machines to overcome local minima.  Shown are thermal 
excitations using in classical thermal annealing, quantum tunneling in quantum annealing, the minimum gain principle in coherent Ising machines, and attractors in dynamical system evolution.  
\label{fig1}} 
\end{figure}

To give an example of such an NP-complete problem, consider  MaxCut (Fig. \ref{fig1}b).  Here, one starts with a graph, where some of the vertices are connected via edges.  The aim then is to group the vertices into two types such that the number of edges (i.e. links between the two groups) is as large as possible.  MaxCut is of direct relevance to problems such as \change{circuit design  \cite{barahona1988application ,1270247}, machine learning \cite{wang2013semi}, and computer vision \cite{collins2004graph, arora2010efficient}}, and therefore even without any mapping is an important problem in its own right.   In a brute force solution of this problem, one requires checking every one of the exponential ways the vertices can be grouped, and finding the largest number of edges. The MaxCut problem can be recast in a physics language as a spin glass problem (Fig. \ref{fig1}a).   To do this, put a binary valued spin on each vertex $ \sigma_i \in \pm 1$, and assign a positive interaction constant $ J_{ij} = 1 $ between the connected vertices and 0 otherwise. The value of spin then encodes which group a vertex is in, and lowers the overall energy for connected spins if they are in different groups \change{$ \sigma_i \sigma_j = -1  $.} This can be written as an Ising Hamiltonian 
\begin{align}
H_P = \sum_{i,j=1}^N J_{ij} \sigma_i \sigma_j +  \sum_{i=1}^N h_{i} \sigma_i  , 
\label{isingham}
\end{align}
where $ N $ is the number of vertices or spins.  We have also included a linear $ h_i $ term for generality, although for MaxCut it is not required.  Finding the minimum energy of (\ref{isingham}) is then equivalent to solving MaxCut. \change{We note that the Ising Hamiltonian (\ref{isingham}) can be related to a Quadratic Unconstrained Binary Optimization (QUBO) problem under a simple change of variables $ \sigma_i = 1 - 2 x_i $, $ x_i \in \{0,1\} $, and hence they can be regarded as equivalent problems. }

The most common way that such optimization problems are solved are on large-scale high performance classical computers, using variants of Monte Carlo methods.  With the demise of Moore's law it is of interest whether alternative methods --- perhaps based on unconventional methods of computing --- could be used to solve such optimization problems.  Alternatives to Turing's concept of a deterministic digital computing machine \cite{turing1937computable} have a long history of investigation, in particular, the analog computers used for  physical simulators to investigate complex problems \cite{bournez2018handbook}. In an analog computer, the computation is performed using coupled physical systems which evolve continuously according to their physical dynamics, implemented by analog electronics or mechanical systems, for example.  These were used predominantly in the first half of 20th century when digital computing speeds were insufficient, and continued to be used for several decades after for specialized applications such as flight simulation, although even these applications have been now rendered obsolete.

The recent interest in the field of quantum simulation  \cite{feynman1982simulating,buluta2009quantum,georgescu2014quantum} in many ways mirrors this development of classical analog computers, and led to a resurgence of interest in realizing analog Ising model simulators. Quantum simulation was in fact one of the early motivations for realizing a quantum computer, based on Feynman's conjecture that a quantum computer should be able to simulate quantum systems more efficiently than classical computers \cite{feynman1982simulating} ---  proven true later by Lloyd \cite{lloyd1996universal}.  However, a large-scale, fault-tolerant quantum computer is still a challenging goal, technologically.  On the other hand, advances in the manipulations of many-body quantum systems using cold atoms, ions, or artificial qubits potentially allow for a way of simulating complex quantum systems without requiring the full \change{ controllability of a quantum computer  \cite{greiner2002quantum,buluta2009quantum,georgescu2014quantum,
labuhn2016tunable,bernien2017probing,keesling2019quantum,scholl2021quantum}.}  This led to the idea that Ising models might be realizable using a quantum simulation approach  \cite{farhi2001quantum,byrnes2011accelerated}, where alternative models of computation could be utilized in order to find the ground state more efficiently. \change{The first large-scale physical implementation of the quantum approach was produced by D-wave Systems, where a 128 qubit quantum annealer was realized, followed by larger scale systems \cite{king2018observation,harris2018phase}.  
Today, there are numerous approaches, incorporating a variety of different techniques (both classical and quantum), which will be described and compared in this review.}

In the classical realm, one of the main drawbacks of analog computers in comparison to digital computers is that they are more susceptible to error, due to the analog storage of the information.  Nevertheless, analog computers can have several advantages over digital computers.  First, the operation of the analog computer is typically highly parallelized.  For a system consisting of many coupled systems which encode the information, each of the systems evolves in parallel, in contrast to digital computers where parallelization is performed across multiple processors. \change{Second, there is no additional overhead due to the implementation of digital logic. In many cases the time evolution of a physical system is continuous, but is discretized and evolved in a step-wise sequence, which requires additional resources not required in analog simulation. }
Analog computing is in many \change{ways} analogous to the way that the brain operates:  there is no predefined algorithm, and its operation is inherently massively parallel \change{and asynchronous}.  This ``natural computing'' approach has intrigued researchers for decades, from both of the point of view of improvements over current computing, as well as understanding the way that biological systems compute.  

In this review, we survey various hardware devices that have been developed with the aim of solving the Ising model; here we call such devices ``Ising machines''.  An important caveat is what exactly we mean by \textit{solve} the Ising model. 
In many applications \textit{suboptimal but still good} solutions are acceptable in practice, hence we shall consider primarily heuristic and approximate solvers.  We focus on discussing their underlying operating principles \cite{vadlamani2020physics}, and introduce the types of technologies that have been used to implement them.  This includes variations of classical thermal annealers, quantum annealers, as well as dynamical system-based solvers including the coherent Ising machine, which have attracted interest recently. Other types of novel computing devices such as those based on hybrid quantum-classical systems are also described. We discuss the performance of the investigated devices, focusing on the scaling with regard to the size of the Ising problem.

\section*{Operating Principles of Ising Machines}

\subsection*{Classical thermal annealing}

One of the fundamental concepts that is encountered in connection to solving the Ising model --- and optimization problems in general --- is annealing. Inspired by concepts in statistical mechanics, the configurations corresponding to the lowest values of a cost function (i.e. energy) are found by gradually lowering the effective temperature of a system.  The basic observation is that at thermal equilibrium, a classical physical system follows statistics according to a Boltzmann (or Gibbs) distribution
\begin{align}
    p_n = \frac{\exp( - \frac{E_n}{k_B T})}{Z} ,
    \label{boltzmanndist}
\end{align}
where $ Z = \sum_n  \exp( - \frac{E_n}{k_B T}) $, $ k_B $ is the Boltzmann constant, $T$ is thermodynamic temperature, and \change{$E_n$ is the energy of} each spin configuration of the Ising model (\ref{isingham}) that we have labeled by $ n = \{ \sigma_1, \dots, \sigma_N \} $. There are $ 2^N $ different energy configurations labeled by $ n $. \change{The lowest energy states appear with higher probability, and the probability of obtaining the ground state, i.e. the desired solution of the Ising model, increases as the temperature is  lowered.}
To produce such a state at thermal equilibrium, the system evolves according to a master equation typically of the form
\begin{align}
    \frac{d p_n}{dt}= - w_{nm} p_n + w_{mn} p_m,
    \label{master}
\end{align}
where rates $ w_{nm} $ are the rates for transition from the $n$th to the $m$th energy state.  The rates are taken such that in the limit of $ t \rightarrow \infty$ the probability distribution follows (\ref{boltzmanndist}).  If  (\ref{master}) is evolved long enough, one is guaranteed to obtain the low energy solutions for a sufficiently low temperature.  The main problem is that particular energy landscapes require extremely long times before thermal equilibrium is reached, due to the possibility of getting trapped in a local minima (Fig. \ref{fig1}c).  The solution to this problem is to gradually lower the temperature, or {\it anneal} the system such that at each temperature the system has a chance \change{to} equilibrate.  Geman and Geman showed that by reducing the temperature with an inverse logarithmic dependence with time, one is guaranteed to obtain the ground state \cite{geman1984stochastic}.  

\begin{figure}[t]
\begin{center}
\includegraphics[width=0.9\linewidth]{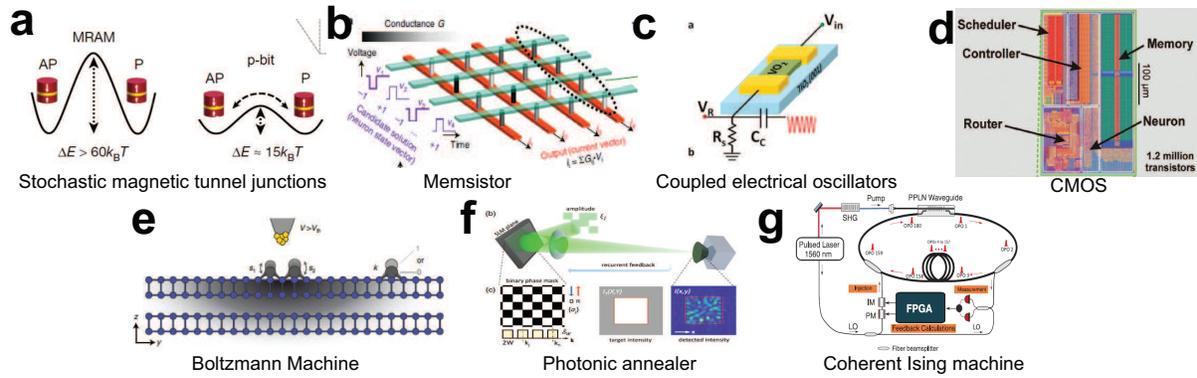} 
\end{center}
\caption{ {\bf Example technologies used to realize various types of Ising machines.} (a) Stochastic magnetic tunnel junction showing difference between conventional MRAM and the probabilistic bit from Ref. \cite{borders2019integer}.  P is the parallel and AP is the antiparallel orientations of the magnetic layers.  (b) Memristor crossbar array to perform matrix-vector multiplication from Ref. \cite{cai2020power}.     (c) Metal insulator VO$_2$ system to realize coupled electrical oscillators from Ref. \cite{shukla2014synchronized}.  (d) 28 nm CMOS chip to realize a 1 million spin Boltzmann machine from Ref. \cite{merolla2014million}.   (e) Co atoms on surface of black phosphorus interacting with a scanning tunneling microscope to realize a Boltzmann machine from Ref. \cite{kiraly2021atomic}.  (f) Spatial light modulator based photonic annealer from Ref. \cite{pierangeli2019large}.   (g) Coherent Ising machine measurement-feedback loop from Ref. \cite{mcmahon2016fully}.   
\label{fig2}} 
\end{figure}
As a classical computer algorithm, simulated annealing \change{(SA)} remains one of the most popular algorithms that can be applied to optimization problems.  Run on a classical digital computer, it is  preferable to perform an equivalent stochastic sampling approach, rather than run (\ref{master}) directly, due to the exponential resources required.  As such, typically one uses various Monte Carlo algorithms employing the Metropolis-Hastings algorithm \cite{metropolis1953equation,hastings1970monte}, such that the desired Boltzmann distribution is obtained.  
More sophisticated classical algorithms are known to provide substantial improvement over \change{SA} such as parallel tempering \cite{PhysRevLett.57.2607,earl2005parallel}, population annealing \cite{wang2015population}, and  isoenergetic cluster moves \cite{zhu2015efficient} to mention a few. For both  parallel tempering and population annealing, multiple copies of the system are prepared in random  initial states. For parallel tempering, each copy has  a different temperature parameter. The temperature is increased for the copies that perform poorly and is decreased for the ones perform successfully. In population annealing,   poorly performing copies are probabilistically removed and those which perform successfully are replicated, while reducing the temperature \cite{kirkpatrick1983optimization,vcerny1985thermodynamical}.

Simulated annealing is most commonly implemented algorithmically in conventional programmable digital classical computers, but has also been implemented in dedicated hardware using digital hardware accelerators and analog natural computing approaches.   
When implemented on dedicated hardware, this provides the chance to exploit the parallelization of digital hardware accelerators and analog computing. For the analog computation approach, numerous physical implementations of Ising and related models have been realized or proposed, including \change{magnetic devices  \cite{camsari2017stochastic,camsari2019p,borders2019integer,sutton2017intrinsic,shim2017ising,arnalds2016new,bhanja2016non,lee2021thermodynamic,mizushima2017large}}, optics \cite{pierangeli2019large,pierangeli2020adiabatic,roques2020heuristic}, memristors \cite{cai2020power,bojnordi2016memristive}, spin-switches \cite{behin2016building}, quantum dots \cite{sarkar2005synthesizing}, single atoms \cite{kiraly2021atomic}, microdroplets \cite{guo2021molecular}, and Bose-Einstein condensates \cite{byrnes2011accelerated,byrnes2013neural} (see Fig. \ref{fig2}).  For example, Camsari, Datta and co-workers have demonstrated that stochastic magnet tunnel junctions can act as probabilistic bits, fluctuating due to thermal energy between either parallel or antiparallel mutual orientations of magnetic domains \cite{camsari2017stochastic,camsari2019p,sutton2017intrinsic} (Fig.\ref{fig2}a).  These are mutually coupled to realize an arbitrary Ising interaction by measuring the orientation of the bits and adjusting the barrier energy between the two orientations.  This was used to factor integers via an adiabatic procedure \cite{borders2019integer}. In another approach, Strachan and co-workers used memristors to perform an analog matrix multiplication of the Ising matrix to evaluate the energy; utilizing the intrinsic hardware noise, they performed a highly parallelized implementation of a Ising model annealer  \cite{cai2020power} (Fig.\ref{fig2}b). For optical systems, the Ising model was realized by encoding the spins using the phase of the light, and a recurrent feedback network was used to produce the Ising couplings \cite{pierangeli2019large,pierangeli2020adiabatic,roques2020heuristic} (Fig.\ref{fig2}f). This was shown to converge towards the Boltzmann distribution (\ref{boltzmanndist}), with the primary advantage being the fast parallelized spin updates. 

For digital-electronic approaches, hardware accelerators using CMOS application-specific integrated circuits \change{(ASICs)} \cite{yamaoka201520k,matsubara2020digital,aramon2019physics,su2020cimspin,yamamoto2020statica} and field-programmable gate arrays (FPGAs) \cite{tsukamoto2017accelerator,matsubara2017ising,yamamoto2017time,patel2020ising,aadit2021massively} have been investigated to solve the Ising model \change{as a type of domain-specific computing}. For example, Yamamoka and co-workers at Hitachi used CMOS circuits to implement $2 \times 10^{4}$ Ising spins, where each spin interacts with up to 5 local spins \cite{yamaoka201520k}.  Here, random thermal effects were introduced by either introducing random spin-flips during calculation of spin values, or applying a low supply voltage to the memory cells, which also introduces randomness at the level of the hardware. Matsubara and colleagues at Fujitsu developed a 8192-spin Ising machine with full connectivity that is based on a digital-CMOS-chip implementation of \change{SA}, where spin updates are performed in \change{parallel \cite{matsubara2020digital,tsukamoto2017accelerator,matsubara2017ising,aramon2019physics}.}  The parallelization allows for a large speedup in comparison to a serial implementation of \change{SA}. We note that in the context of machine-learning accelerators\cite{reuther2019survey}, hardware implementations of Boltzmann machines have been investigated with CMOS ASICs \cite{arima1991336,alspector1991relaxation,merolla2014million} (Fig. \ref{fig2}d), FPGAs \cite{skubiszewski1992extact,zhu2003fpga,kim2009highly,kim2010large,le2010high,kim2014fully}, and GPUs \cite{ly2008neural,zhu2013large}. 
The similarity of the underlying energy model of Boltzmann-machine hardware accelerators suggests that such technologies could be adapted to act as Ising solvers.  Recently, Patel, Salahuddin and co-workers demonstrated an FPGA implementation of the restricted Boltzmann machine's stochastic sampling algorithm to solve the Ising problem \cite{patel2020ising}. \change{Here the problem is mapped to a bipartite version and each group of spins is updated applying parallel SA \cite{PhysRevE.100.012111,yamamoto2020statica}}. The inherent parallelism of this architecture allows parallel sampling which provides substantial improvement over \change{SA}.

\subsection*{Dynamical system solvers}

In a thermal annealer, at any given point of time during the evolution, the system is ideally in a state that is at thermal equilibrium, following the Boltzmann distribution. Likewise, \change{as we shall see in the next section, } in a quantum annealer the system ideally remains in the ground state of the instantaneous Hamiltonian. To ensure these conditions, annealing must proceed sufficiently slowly such as to maximize the probability that the minimum energy state of the Ising model is obtained. In contrast to such annealing-based approaches, alternative strategies where the system evolution is much faster than thermal equilibriation and adiabatic timescales also exist. In such dynamical system approaches, the state of the system is driven towards the lowest energy state of the Ising model.  An early example of such a dynamical solver was proposed by Hopfield and collaborators  \cite{hopfield1982neural,hopfield1985neural} where electronic circuits were used to realize (\ref{isingham}). In this section, we explain three types of dynamical system solvers: coupled oscillators, coherent Ising machines, and chaotic systems.


\subsubsection*{Oscillator-based computing}
In the 1950s, von Neumann and Goto pioneered a type of analog computer based on coupled oscillators called the ``parametron'' computer \cite{von1957non,wigington1959new,goto1959parametron}.  Here the state information, such as the configuration of an Ising spin, is represented by the phase of an oscillator. In the presence of a non-linearity, an oscillator with resonant frequency $ \omega_0 $ can be phase locked with a pump frequency $ 2 \omega_0 $ with two possible stable phases, $ 0 $ or $ \pi$ that represent the digital information.  
In its original conception, information processing in the parametron computer occurs as a sequence of logical gates.  However, it was also shown that a computation can be performed in a more parallel, natural computing approach (see Refs.  \cite{csaba2020coupled,raychowdhury2018computing} for a review). Such coupled oscillators can be used for solving combinatorial optimization problems, such as the Ising model.
The basic idea of oscillator based computing can be captured by the Kuramoto model. This describes a system of oscillators mutually coupled by an interaction \cite{kuramoto1975international,acebron2005kuramoto,breakspear2010generative}.  Consider $ N $ oscillators, labelled by the index $ i $ that oscillate with frequency $ \omega_0 $.  Denote the phase of the $i$th oscillator by $ \phi_i $.  Mutually coupling the 
oscillators, the dynamical system can be described by \change{
\begin{align}
\frac{d \phi_i}{dt} = \omega_0 + K \sum_{j} J_{ij} \sin ( \phi_i - \phi_j) + K h_i \sin ( \phi_i - \omega_0 t ) ,
\label{kuramoto}
\end{align}
where $K $ is a coupling parameter which controls the overall contribution of the Ising dynamics.  
In the rotating frame,  the $ \sin ( \phi_i - \phi_j) $ factor has two steady-state solutions $ \frac{d \phi_i}{dt} = 0 $, where the phases are either in or out of phase. The $  \sin ( \phi_i  - \omega_0 t )  $ term is also stable when $ \phi_i - \omega_0 t = 0,\pi$. } This means that the dynamical set of equations will converge to a particular configuration of phases, from which a spin readout can be performed.  For the constant  $ J_{ij} $ and $ h_i = 0 $ case the Kuramoto model can be analytically solved to show a dynamical phase transition between unsynchronized oscillators to an synchronized one for particular interaction strengths. This type of dynamics has been applied to numerous types of applications in artificial intelligence problems, such as image processing, pattern recognition and generation  \cite{csaba2020coupled,raychowdhury2018computing}.  

Wu, Chua and co-workers proposed using such oscillators to solve the graph coloring problem using a system of coupled LC circuits \cite{wu1995application,wu1998graph}.
Here the aim is to color the vertices of a graph with $ k $ colors such that no adjacent vertices have the same color.  For $ k\ge 3$ this is a NP-complete problem; in Ref. \cite{wu1995application} it was found that for $ k = 2$ the scheme is able to correctly find solutions, but for the more difficult $ k = 3$ case the scheme only succeeded for a subset of problem instances.  This approach was theoretically further developed, 
\change{including explicitly extending to the case of solving Ising problems and analyzing various possible physical implementations\cite{wu2011clustering,kalinin2018global,wang2019oim,afoakwa2021brim,mcgoldrick2021ising,albertsson2021ultrafast}.} Oscillator networks have been experimentally demonstrated with various systems such as bulk analog electronic \cite{wang2019oim,chou2019analog,xiao2019optoelectronics,saito2020amoeba}, 
$ \text{VO}_2$ insulator-to-metal transition  \cite{shukla2014synchronized,parihar2017vertex,dutta2020ising} (Fig. \ref{fig2}c)\change{, spin\cite{zahedinejad2020two,houshang2020spin}} and integrated CMOS electronic \cite{mallick2020using,bashar2020experimental,ahmed2021probabilistic} oscillators to solve problems such as graph coloring, maximum independent set, and the Ising model. In several of these studies, by adding noise and turning on the interactions smoothly, it was found that this enabled the network to find low-energy solutions of the Ising model. 

\change{A related approach called memcomputing uses networks of Boolean logic gates as a  dynamical system solver.  These have conceptual similarities with oscillator-based Ising machines even if they are not explicitly constructed from networks of oscillators \cite{traversa2015universal,diventra2018perspective}. Such approaches have been applied successfully to frustrated-loop Ising model instances \cite{sheldon2019taming,aiken2020memcomputing}.    }

\subsubsection*{Coherent Ising Machine}

Yamamoto and colleagues have investigated a particular class of oscillator-based Ising machines, dubbed ``coherent Ising machines'' (CIMs), that are naturally suited to being implemented with {\it optical} oscillators\cite{utsunomiya2011mapping,wang2013coherent,marandi2014network,mcmahon2016fully,inagaki2016coherent,yamamoto2017coherent,leleu2019destabilization,hamerly2019experimental,yamamoto2020coherent,honjo2021spin} (Fig. \ref{fig2}g). Each Ising spin $\sigma_i$ in a CIM is encoded in the \textit{phase} $\phi_i$ of light in an optical mode. To enforce binary spin values $\phi_i = 0,\pi$, CIMs use \textit{degenerate optical parametric oscillators} (DOPOs), which are a form of parametric oscillator in which phase-sensitive gain yields oscillations either in-phase or out-of-phase with respect to the oscillator's pump light \cite{wang2013coherent,marandi2012all}. Each DOPO represents a single spin and is part of a network of DOPOs that are coupled together such that the coupling between a pair of DOPOs 
is proportional to the Ising spin-spin coupling $J_{ij}$. Several ways to realize couplings have been proposed\cite{byrnes2011accelerated,wang2013coherent,marandi2014network,mcmahon2016fully,inagaki2016coherent}, but for the experimental demonstrations performed thus far, the details of the coupling scheme are not crucial for understanding the Ising-solving capability of each CIM implementation. If one models DOPOs as \textit{classical} oscillators, then the time evolution of a CIM can be modeled by the following system of coupled differential equations in the rotating frame:\cite{wang2013coherent,yamamoto2017coherent}
\begin{align}
\frac{d a_i}{dt} = -\gamma a_i + r a_i^* -\kappa | a_i |^2 a_i - g \sum_{j} J_{ij} a_j - g h_i + n_i,
\label{eq:conttime_CIM}
\end{align}
where each $ a_i $ is a complex number representing the optical field in the $i$th mode, $ \gamma $ is the decay rate of the photons from each mode, $r$ is the amplification provided by OPO gain, $\kappa$ is the coefficient of nonlinear loss due to OPO gain saturation, \change{$g$ is a coupling constant determining the strength of the Ising interactions}, and $ n_i $ are Langevin noise operators associated with the photon decay and nonlinear gain. The effect of the Ising terms $-g \sum_{j} J_{ij} a_j - g h_i$ can be thought of as additional loss terms that act on the $i$th mode. 

The key new elements in these equations of motion relative to those for Kuramoto oscillators (\ref{kuramoto}) are that the oscillator amplitudes are also explicitly considered, in addition to their phases, and there are now loss and gain terms; it is these terms that are responsible for DOPOs having an oscillation threshold. If one considers a single DOPO, the loss and gain terms result in the DOPO being bistable: above threshold, a DOPO will oscillate either exactly in-phase or exactly out-of-phase. Since the Ising terms can be interpreted as spin-configuration-dependent loss, one can interpret the DOPO network as having a collective-oscillation threshold that is lowest when the Ising terms are smallest, and hence when the represented spin configuration has minimum energy. If the CIM is operated in a way where the gain $r$ is slowly increased from $0$ (when the DOPO network is below threshold) to ever-higher values (i.e., $r$ is not a constant, but rather a monotonically increasing function of time $t$), then, in the absence of noise $n_i$, the DOPO configuration with the lowest loss should oscillate first (see the minimum gain principle illustrated in Fig.~\ref{fig1}c) and the solution to the Ising problem can be read out by measuring the phases of the light from each DOPO. An important point to note is how slowly $r$ can be increased and have the CIM still oscillate in the ground state for a length of time sufficient to allow measurement: even in the complete absence of noise (which is not experimentally realistic, but can be programmed in a computer simulation), the CIM does not find the exact solution to arbitrary Ising problems in polynomial time. \change{An important technicality that arises in the CIM model (\ref{eq:conttime_CIM}), as well as in other oscillator-based Ising machines where the oscillators have both amplitude and phase degrees of freedom (as opposed to just phase), is that if the amplitudes $\left| a_i \right|$ of the oscillators are not equal, then the system will tend to minimize the energy of an Ising instance with a different $J_{ij}$ matrix than the desired one. This phenomenon is sometimes referred to as a \textit{broken mapping due to amplitude heterogeneity}. An intuitive fix is to add a feedback mechanism that forces the amplitudes $\left| a_i \right|$ to be equal; this has been studied for XY machines\cite{kalinin2018networks} and Ising machines\cite{leleu2019destabilization}.}

The classical description of a CIM (\ref{eq:conttime_CIM}) is sufficient to explain the results obtained in the experimental demonstrations\cite{marandi2014network,takata201616,inagaki2016large,mcmahon2016fully,inagaki2016coherent,hamerly2019experimental,okawachi2020demonstration} to date, since these experiments have used DOPOs with fairly large roundtrip (photon) loss. However, with sufficiently low loss, each DOPO can generate an appreciable amount of quadrature squeezing, and in this regime the CIM's dynamics are more faithfully modeled quantum mechanically \cite{yamamoto2017coherent}. 
An interpretation for CIM operation that arises in the quantum-mechanical formulation is that each DOPO begins in a squeezed state that is approximable by a coherent superposition of in-phase and out-of-phase coherent states $| \alpha \rangle + |-\alpha \rangle$, so the below-threshold state of the CIM is one in which every spin configuration is represented in superposition, and when the CIM goes through threshold, one of the configurations is selected. 
It is an open question to what extent quantumness of the DOPO network may improve (or impair) the computational performance of a CIM \cite{yamamoto2020coherent}. A quantum model for a machine conceptually quite similar to a quantum-regime CIM, in which superpositions $| \alpha \rangle + |-\alpha \rangle$ are also formed, has been studied by Goto\cite{goto2016bifurcation}; Goto found that the machine acts as an adiabatic quantum computer when the pump rate (the equivalent of $r$ in the CIM model above) is increased from 0 sufficiently slowly. This theoretical connection suggests that insights into the solution mechanisms of quantum annealers might be helpful for understanding CIMs, \change{especially CIMs in which the coupling between OPOs is conservative rather than dissipative,} and vice versa.

Besides the CIM, there have been proposals and demonstrations of several different types of optical and optoelectronic Ising and Ising-like machines in addition to the ones already cited in the subsection on thermal annealers: systems based on coupled lasers\cite{tamate2016simulating,babaeian2019single,parto2020realizing}, optoelectronics\cite{bohm2019poor}, exciton-polaritons\cite{lagoudakis2017polariton,berloff2017realizing,kalinin2018global,kalinin2018simulating,kyriienko2019probabilistic}, and electromechanical systems \cite{mahboob2016electromechanical, tezak2019integrated}. 

\subsubsection*{Chaos in dynamical-system solvers}

In an ergodic system, the dynamics are such that the system visits 
all parts of configuration space.  
This is an attractive idea in the context of solving the Ising model, since in many 
approaches getting trapping in local minima is the cause of the exponential slowdown. 
Numerical studies studying thermal relaxation have showed that the process is strongly
non-ergodic, and do not visit all parts of configurational space \cite{bernaschi2020strong}. 
Several studies have suggested by modifying the dynamics to include chaos, 
this can lead to an improvement in performance \cite{ercsey2011optimization,molnar2018continuous,leleu2019destabilization,leleu2020chaotic}. 
Ercsey-Ravasz and Toroczkai proposed and studied\cite{ercsey2011optimization} a limit-cycle-free dynamical system whose fixed-point attractors are the solutions of a given optimization problem. The particular optimization problem they designed their system for was $k$-SAT, which is---like the Ising problem---an NP-complete decision problem with an NP-hard optimization version. The formulation of the dynamical system involved both state variables $s_i$ corresponding to the variables in the $k$-SAT problem (analogous to spin variables for an Ising problem) and auxiliary variables $ a_i $. The theoretical property of the dynamical system that it avoids becoming stuck in local minima of the $k$-SAT cost function is very appealing. However, this comes at a price: the auxiliary variables $ a_i $ grow exponentially in time. Two interpretations or implications of this are as follows: an analog hardware implementation of the dynamical system will require an exponentially growing amount of energy to operate (a prototype CMOS demonstration\cite{yin2017efficient} for problems with up to 50 variables artificially capped the signals representing the auxiliary variables at 1 V), and a digital hardware implementation that integrates the differential equations will need to use exponentially small timesteps because the differential equations become stiff (observed in Refs.~\cite{ercsey2011optimization,molnar2018continuous}).

It was found empirically in Ref.~\cite{ercsey2011optimization}, through numerical simulations, that the dynamical system undergoes a transient period of chaos while solving difficult instances of the $k$-SAT problem, but not when solving easy instances. It was suggested in this work that chaos might be unavoidable in approaches to solving hard optimization problems.
A discrete-map optimization algorithm\cite{elser2007searching} applied to solving both $k$-SAT and Ising problems was also found to exhibit chaotic dynamics.
The general approach in Ref.~\cite{ercsey2011optimization} for designing a limit-cycle-free dynamical system that avoids being trapped in $k$-SAT local minima through the use of auxiliary variables has been adopted for Ising solving\cite{leleu2019destabilization}, and has been implemented and benchmarked with an FPGA\cite{leleu2020chaotic}.

\subsection*{Quantum approaches}

\subsubsection*{Quantum annealing}
Quantum annealing (QA) \cite{RevModPhys.90.015002,RevModPhys.80.1061,hauke2020perspectives} is a heuristic algorithm based on the quantum adiabatic theorem and was first proposed by Apolloni et al. \cite{apolloni1989quantum} \change{and studied in the context of the Ising model by Kadowaki and Nishimori \cite{PhysRevE.58.5355}.}  In this algorithm, the system is initially prepared in the ground state of a Hamiltonian $H_0$ where its ground state is known.  A common choice for this initial Hamiltonian is 
\begin{align}
H_{0}=-\sum_{i=1}^{N} \sigma_{i}^{x},
\label{simpleham}
\end{align}
where $N$ denotes the number of qubits, and the ground state is the uniform superposition of all possible configurations $|+\rangle^{\otimes N}$, \change{where $|+\rangle=(|0\rangle+|1\rangle)/\sqrt{2}$.} The Hamiltonian is  gradually reweighted to the desired problem Hamiltonian $H_P$ according to
\begin{align}
H =(1- \lambda (t)) H_0+ \lambda (t) H_P ,
\label{Isingham}
\end{align} 
where $\lambda (t) \in [0,1] $ is the annealing schedule. 
The annealing process can be viewed as $H_0$ introducing quantum fluctuations \change{originating from the non-commutability of  $H_P$ and $H_0$}.  These fluctuations are gradually reduced to reach the low-energy configuration of the classical energy function $H_P$. Based on the quantum adiabatic theorem, if one performs the sweep sufficiently slowly, the system remains in its instantaneous ground state throughout the evolution \cite{farhi2000quantum,RevModPhys.90.015002}.  The sweep time for which  the adiabaticity can be achieved is proportional to a negative power of the minimum energy gap between two lowest energy levels during the sweep \cite{roland2002quantum,amin2008effect,schaller2006general,lidar2009adiabatic}. The use of quantum fluctuations in QA has been hypothesized as a potential resource for a speedup over classical methods. Quantum tunneling allows the system to pass through energy barriers which have a higher energy than available in the state (see Fig. \ref{fig1}c). 
However, despite several decades of investigation, the computational role of coherent tunneling in providing speedup still is not completely understood \cite{RevModPhys.90.015002,katzgraber2015seeking,PhysRevX.6.031010}. Part of the reason for this is the difficulty of simulating QA on classical computers due to the large computational overhead. The only quantum hardware that has been able to directly test QA with a large number of qubits to date is that developed by D-Wave Systems.  While this technology still suffers from limitations such as the presence of decoherence, control errors, and limited connectivity, several studies have shown that quantum effects do play a role in the D-Wave machine \cite{albash2018demonstration,denchev2016computational,boixo2016computational}.   For problem instances that possess tunneling barriers, QA and quantum-inspired classical algorithms that mimic tunneling \cite{albash2018demonstration} have been shown to have an advantage over \change{SA}.

\subsubsection*{Hybrid Quantum-Classical Algorithms}

The aim of variational quantum algorithms  \cite{cerezo2020variational} is to solve classical and quantum optimization problems by combining a parametrized quantum circuit with a classical optimizer to obtain the variational parameters.  The parametrized quantum circuit can be thought of as preparing a variational quantum state, which is optimized to give the lowest energy state of a given Hamiltonian.  These algorithms are believed to be strong candidates to achieve a practical quantum advantage on noisy intermediate scale quantum (NISQ) devices \cite{preskill2018quantum}. 
In the context of combinatorial optimization problems, the quantum approximate optimization algorithm (QAOA) \cite{farhi2014quantum} has particularly attracted a lot of interest, partially as a result of the existence of theoretical guarantees on the approximation ratio that it can achieve for certain classes of optimization problems \cite{farhi2014quantum,farhi2014quantumE3LIN2}.

The QAOA algorithm can be viewed as a Trotterized version of QA with a parametrized annealing pathway \cite{zhou2020quantum}. 
The system is initially prepared in $|+\rangle^{\otimes N}$, the ground state of the  Hamiltonian (\ref{simpleham}).  The parameterized quantum circuit transfers the initial state to the ground state of the target problem Hamiltonian \change{(in the ideal case)} by alternately applying the unitary operator corresponding to the problem Hamiltonian $e^{-i \gamma_{j} H_P}$ and the unitary operator $e^{-i \beta_{j} H_{0}}$.  This sequence generates the following quantum variational state 
 \begin{equation}
     |\psi(\boldsymbol{\beta}, \boldsymbol{\gamma})\rangle=e^{-i \beta_{p} H_{0}} e^{-i \gamma_{p} H_P} \cdots e^{-i \beta_{1} H_{0}} e^{-i \gamma_{1} H_P}|+\rangle^{\otimes N} ,
 \end{equation}
where $ \boldsymbol{\gamma}=\left(\gamma_{1}, \gamma_{2}, \cdots, \gamma_{p}\right) \in[0, 2\pi]^p$ and $\boldsymbol{\beta}=\left(\beta_{1}, \beta_{2}, \cdots, \beta_{p}\right) \in[0, \pi]^p$ are $2p$ variational parameters and $p$ determines the circuit depth. Next, a classical optimizer is applied to find the optimal $ \boldsymbol{\beta}, \boldsymbol{\gamma} $ that optimizes the energy expectation $
    E(\boldsymbol{\beta}, \boldsymbol{\gamma})=\left\langle\psi(\boldsymbol{\beta}, \boldsymbol{\gamma})\left|H_P\right| \psi(\boldsymbol{\beta}, \boldsymbol{\gamma})\right\rangle $
by updating the variational parameters iteratively.  Various approaches have been applied for this classical optimization step such as brute force grid search \cite{farhi2014quantum}, gradient descent methods \cite{guerreschi2017practical}, and machine learning models \cite{khairy2020learning}. 
A key feature of QAOA is that the computational power increases with $p$ \cite{zhou2020quantum,pagano2020quantum} in contrast with QA where the performance does not always improve with annealing time \cite{farhi2014quantum}.   It is has been found that under reasonable complexity-theoretic assumptions, QAOA with $p=1$  can not be efficiently simulated with classical computers  \cite{farhi2016quantum}, or it implies that P=NP. This has led to speculations that QAOA may be able to demonstrate \change{a quantum computational advantage} in the context of an optimization problems on near term quantum computers \cite{farhi2016quantum}. However, the class of problems that can be solved efficiently with shallow circuits may not be representative for problems of practical interest. For example, for all-to-all connected Ising models and MaxCut, it has been shown that deep circuits may be required \cite{harrigan2021quantum}.  Therefore, to benchmark computational  advantages of QAOA  against classical algorithms one needs to go far beyond problems that can be solved with a shallow circuit and explore power of QAOA at  intermediate depths. However, at current technological levels such circuits are prone to decoherence and gate errors \cite{harrigan2021quantum}.

QAOA has been demonstrated at the small-scale on platforms such as superconducting qubits \cite{harrigan2021quantum}, photonics \cite{qiang2018large} to trapped ions \cite{pagano2020quantum}.  To date, no large-scale (i.e.,  $ N > 50 $) demonstrations of QAOA have been experimentally performed. 
We note that classical simulations showing expectation results for single-layer ($p=1$) QAOA on problems with up to $N=10^5$ spins have been performed\cite{ozaeta2020expectation}, but as is the case with quantum annealers, it is expected that large-scale quantum hardware will be needed to properly evaluate the performance of QAOA in general.

\subsubsection*{Other quantum algorithms}

Several other quantum algorithms have been proposed to solve combinatorial optimization problems (see Ref. \cite{sanders2020compilation} for a review).  These include using amplitude amplification based approaches, and quantum simulated annealing (not to be confused by simulated quantum annealing below), where the aim is to prepare the quantum Gibbs state, a superposition state with Boltzmann probabilities (\ref{boltzmanndist}) as the amplitudes. Attaining the quantum Gibbs state is done by performing a quantum walk, such that after many iterations the desired coherent Gibbs state is obtained \cite{somma2008quantum,boixo2015fast,lemieux2020efficient}. Then in a similar way to thermal annealing, the temperature is gradually lowered in order to obtain a low-energy state.

\subsection*{Other classical algorithms}

\subsubsection*{Quantum-inspired classical algorithms}

Novel types of classical algorithms have been proposed which are inspired by quantum algorithms.  Such quantum-inspired algorithms are run on conventional computing hardware or on \change{digital} hardware accelerators, hence are classical approaches, but use concepts that originate from quantum mechanics in the algorithm. We briefly summarize several approaches in this direction as below.  

In simulated quantum annealing (SQA), quantum Monte Carlo is applied to estimate the low-energy states of the QA Hamiltonian \cite{bapst2013thermal,das2005quantum,crosson2016simulated}.   To perform the quantum Monte Carlo, a stoquastic QA Hamiltonian is mapped to a classical Hamiltonian by introducing an extra spatial dimension, corresponding to imaginary time. A stoquastic Hamiltonian is characterized by
having only nonpositive off-diagonal elements in the computational basis. The new Hamiltonian has the equivalent equilibrium properties with the original QA Hamiltonian \cite{andriyash2017can}.  The mapping can be implemented both in discrete time by applying the  Trotter-Suzuki decomposition, or in continuous time by applying a path integral \cite{andriyash2017can}. 
Quantum Monte Carlo in the continuous time limit samples the equilibrium thermal state of a quantum system (as opposed to directly simulating its unitary time evolution) and can generate Boltzmann distributed states (\ref{boltzmanndist}).  At sufficiently low temperatures, SQA can mimic tunneling effects. SQA can also generate entangled ground states that occur during the adiabatic evolution. It can thus faithfully predict the  performance of QA for stoquastic Hamiltonians.

Several other quantum-inspired classical algorithms  based on dynamical system evolution have been proposed which we briefly describe.  The simulated coherent Ising machine simulates the CIM on a classical computer and has shown a speedup in comparison to a physical implementation of a CIM applying FPGA \cite{king2018emulating} and GPU \cite{tiunov2019annealing}.   The key observation here is that simulating such behavior is described by a set of coupled equations of the form (\ref{kuramoto}) or (\ref{eq:conttime_CIM}) only scale with the number of spin variables $ N $, rather than the configurational space $ 2^N$.  This makes a simulation of the coupled oscillator system efficient. 

\change{Another approach is simulated bifurcation (SB), which is based on simulating adiabatic evolutions of classical nonlinear Hamiltonian dynamical systems. This algorithm is the classical counterpart of bifurcation-based adiabatic quantum computation \cite{goto2016bifurcation}.} Two branches of the bifurcation in each non-linear oscillator represents two states of each Ising spin. In 2019, Toshiba developed an FPGA and GPU-based SB machine  showing excellent performance due in part to its high parallelizability \cite{goto2019combinatorial,tatsumura2019fpga}. The operational mechanism of SB algorithm is based on an adiabatic and ergodic search.  \change{Later, two other variants of SB  were introduced, called  the ballistic simulated bifurcation algorithm (bSB) and the discrete simulated bifurcation algorithm (dSB) \cite{goto2021high} that far outperform the original SB in terms of both speed and solution accuracy. These new algorithms apply new approaches, such as a quasi-quantum tunneling effect.
Recently, a multi-chip architecture using a partitioned version of the SB algorithm was implemented with FPGAs, showing that large scale Ising problems can be handled using the method \cite{tatsumura2021scaling}.  }
Both CIM simulations and the SB algorithm are highly amenable to parallelization, by simultaneously updating at each time step $N$ coupled-oscillator variables. This is in contrast to SA, which canonically involves sequential updates of spins, with simultaneous updates allowed only for isolated spins.

Yet another quantum-inspired algorithm involves tensor networks which are a powerful framework that provides representations of complex quantum states based on their entanglement structure \cite{orus2019tensor}. Tensor networks have been applied as an ansatz to solve optimization problems \cite{orus2019tensor, alcazar2021enhancing}.  Such an approach was used in the context of dynamic portfolio optimization, which can be encoded as an Ising problem\cite{mugel2020dynamic}.

\subsubsection*{Machine learning approaches}
Machine learning algorithms can be applied to either boost the performance of traditional classical solvers and quantum algorithms \cite{mohseni2021deep} or can work as a stand-alone solver.   For example, some machine learning algorithms have been applied to accelerate Monte Carlo \change{simulations} \cite{bojesen2018policy, PhysRevB.95.035105}. 
It also has been applied as a stand-alone solver and found to have excellent performance in the context of portfolio optimization \cite{alcazar2021enhancing}. The emergence of the methods that are more sample-efficient make them more scalable to large-scale problems.   Deep learning based methods applying reinforcement learning \cite{bello2016neural, dai2017learning}, \change{graph neural networks \cite{zhou2020graph,dwivedi2020benchmarking, schuetz2021combinatorial},} and   neural attention mechanism \cite{vinyals2015pointer} also have been investigated as solvers for combinatorial optimization problems.

\section*{Computational complexity}

One may ask how the various computing approaches discussed in this review relate to computational complexity, and what the prospects are to devise an Ising machine that can solve Ising problems efficiently (i.e., in polynomial time). While the  $\text{P} \stackrel{?}{=} \text{NP} $ question remains an open problem, it is widely conjectured that $ \text{P} \neq \text{NP} $, i.e., that certain problems in NP, including the Ising problem, are fundamentally more difficult to solve than those in P. Indeed, decades of computer science, physics, mathematics, and operations research has failed to find a polynomial-time algorithm that solves any \change{NP-complete} problem. The explosion of interest in quantum computing since the 1990s was kicked off by the discovery that integer factorization could be performed in polynomial time on a quantum computer \cite{shor1999polynomial}.  It is thus conjectured that the BQP complexity class---decision problems a quantum computer can solve in polynomial time with an error probability of at most 1/3---is a larger class than P, i.e., $ \text{P} \subseteq \text{BQP} $.  The associated class for a probabilistic Turing machine, the BPP class, is meanwhile conjectured to be equivalent to P, i.e.,  $ \text{P} =  \text{BPP} $,  \change{which in general remains unproven but is true if a suitable pseudorandom number generator is available \cite{arora2009computational}}. Although there is no proof that quantum or probabilistic computers cannot solve NP-complete problems such as the Ising problem in polynomial time, it is considered unlikely \cite{aaronson2010bqp}.

The complexity-class arguments above concern solving the Ising problem in the sense of being able to find the {\it exact} ground state for all possible problem instances (all possible $J_{ij}, h_i $ in (\ref{isingham})).  However, as mentioned in the introduction, an approximate solution with an energy close to the true ground state is often acceptable for practical applications. 
We may distinguish three approaches for solving combinatorial optimization problems such as the Ising problem: exact algorithms, approximation algorithms, and heuristic algorithms. Exact algorithms are ones designed to find exact solutions in a way that guarantees that the returned solutions are exactly optimal. 
Approximation algorithms return solutions that are not necessarily optimal but provide a guarantee on how far a returned solution is from being optimal. Heuristic algorithms return solutions without any guarantee on the quality of them; because of this lack of theoretical guarantee, the primary basis for trusting an heuristic algorithm is from prior empirical (benchmarking) results. Both approximation and heuristic algorithms tend to be practical to run on large problems.

For approximation algorithms, we may ask what solution quality (formally, \textit{approximation ratio}) one could hope to guarantee with any algorithm. The MaxCut problem, and hence the Ising model, is APX-hard (i.e. approximable-hard). The main consequence of this is that, assuming  $ \text{P} \ne  \text{NP} $, there exists no polynomial-time approximation algorithm for the Ising problem that will guarantee a solution arbitrarily close to the exact solution \cite{papadimitriou1991optimization}. However, there does exist a polynomimal-time approximation algorithm for MaxCut that will find solutions a fixed distance from the optimal solution: the Goemans-Williamson algorithm is guaranteed to find solutions within $ \approx 12 \%$ of the optimal value \cite{goemans1995improved}. It is also known that it is NP-hard to approximate MaxCut with solutions guaranteed to be closer than $ \approx 6 \%$ to the optimal\cite{haastad2001some}, so it is expected (assuming $ \text{P} \ne  \text{NP} $) that no polynomial-time approximation algorithm achieving this approximation closeness will be possible. In many practical settings it is desirable to find solutions to MaxCut problems having distance from the optimal solutions that is better than $ \approx 12 \%$ or even $ \approx 6 \%$, and this motivates the use of heuristic algorithms to solve MaxCut in practice.

Most Ising machines are heuristic solvers---that is, they can be thought of as physical machines realizing heuristic optimization algorithms. As such, they typically do not provide any approximation-ratio guarantees. The potential advantages of Ising machines largely lie outside the realm of complexity theory: there is the possibility that Ising machines have a polynomially improved scaling or constant-prefactor advantage in comparison to existing heuristic algorithms running on conventional processors. i.e., it is generally expected that Ising machines, regardless of their underlying algorithm or practical hardware implementation, will still have runtimes scaling exponentially in order to achieve near-optimal solutions, but that the exponent may be smaller than for a conventional solver, or that the constant factor in front of the exponent may be smaller. A small difference in the exponent can make a large difference in runtimes for large problem sizes, and the fast clock speeds of various physical implementations, which give rise to constant-factor improvements, could lead to significant practical speedups versus conventional state-of-the-art solvers. 

\section*{Computation performance comparisons \label{Performance}}


Since the utility of any Ising machine is in its ability to solve a given Ising problem both quickly and accurately, an important task is to benchmark the performance and compare between competing methods. 
Here we will direct our attention particularly to various Ising solvers that have been experimentally tested for relatively large systems $ N \ge 50 $.   We consider only large systems as it is rather difficult to extract any scaling relation for smaller systems, and 
by choosing experimentally realized systems this will tend to focus on technologies that are relatively near to maturity.  For the figures of merit,  we focus upon two of the most commonly employed quantifiers: the success probability and the time-to-solution.  We first define each of these.

For the success probability, this is defined as the probability that the exact ground state of the Ising model is obtained for a single run of the Ising machine.  
The success probability is a quantity that depends inherently on algorithmic parameters.  For example, for annealing methods under ideal conditions, longer annealing times generally result in higher success probabilities. In this sense, the success probability can also be made arbitrarily close to 1 in the ideal case.  However, other factors typically prohibit approaching unit success probability, due to practical considerations.  For example, in quantum annealers the annealing time must be within the coherence time in order to maintain a quantum superposition. 
In this sense, the success probability still has a meaning, since it often involves a trade-off with practical considerations.  For our comparison of success probabilities, we generally quote the best performing value available in the literature.  

One of the limitations of the success probability as a figure of merit is that it does not take into account of how much time it takes for a single run of the Ising machine.  An Ising machine will typically perform multiple runs when attempting to solve a problem and Ising machines are often optimally operated for a choice of run parameters where the success probability for a single run is not maximized, but where each run takes only a short amount of time. The time-to-solution is another figure of merit that takes into account of the time taken to perform a single run on a given Ising machine, as well as the success probability.  Suppose that $ r $ runs of a particular scheme are performed, which each have a success probability $ p_{\text{suc}} $ of obtaining the ground state .  Then the collective probability of getting at least one successful run, in which the ground state is found, is $ 1 - (1-p_{\text{suc}})^r$.  Now let us demand this collective probability to be a desired value, say 99\%, and each run takes a time $ \tau $.  The time-to-solution is then related to the success probability as
\begin{align}
T_{\text{sol}} = \tau \frac{\ln 0.01}{\ln (1-  p_{\text{suc}} )} .
\end{align}
This measure takes into account of differences in clock speed of various approaches, and also allows for various approaches to choose their optimal parameters such that the best performance of the machine can be extracted. 

\begin{figure}[t]
\begin{center}
\includegraphics[width=\linewidth]{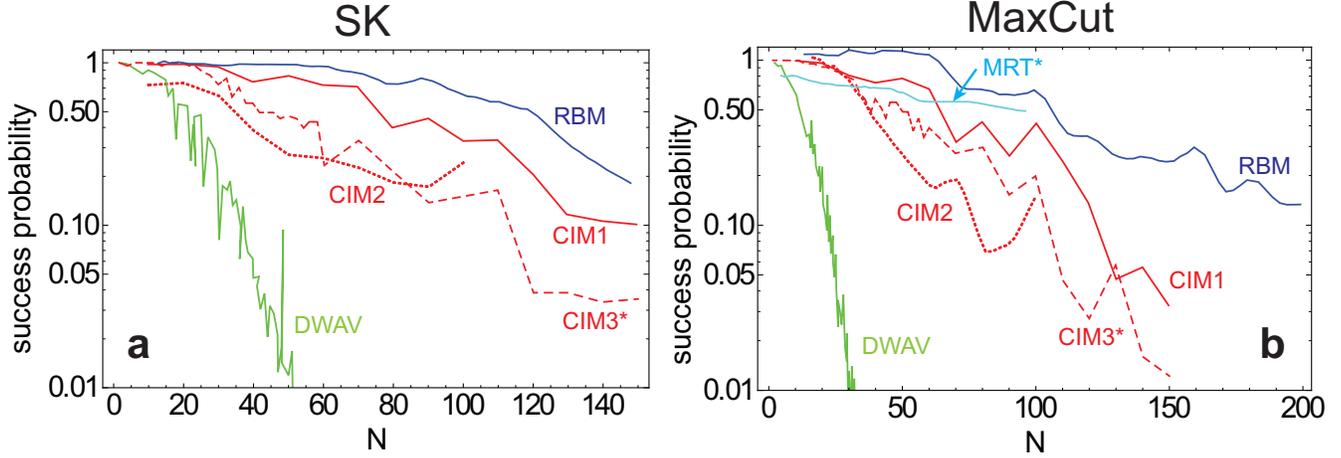} 
\end{center}
\caption{ {\bf Success probability comparison of various Ising machines.} The probability of obtaining the ground state is shown for (a) Sherrington-Kirkpatrick (SK) and (b) dense MaxCut  problems.  For the SK problem, $ J_{ij} $ is chosen from $ \pm 1 $ with equal probability. 
\change{The MaxCut problem is mapped onto the Ising model by setting $ J_{ij} $ to 0 and 1 with equal probability.} In both cases $ h_i =0 $. The labels for each line and their references are given in Table \ref{table1}. \change{Error bars on original data where present have been omitted for clarity.}
Here, CIM1, CIM2, CIM3, DWAV and RBM are benchmarked on the same problem instances. \change{Data reported for theoretical predictions, rather than being directly measured from an hardware implementation, are labeled with *.}
\label{fig4}} 
\end{figure}

\begin{figure}[t]
\begin{center}
\includegraphics[width=0.8\linewidth]{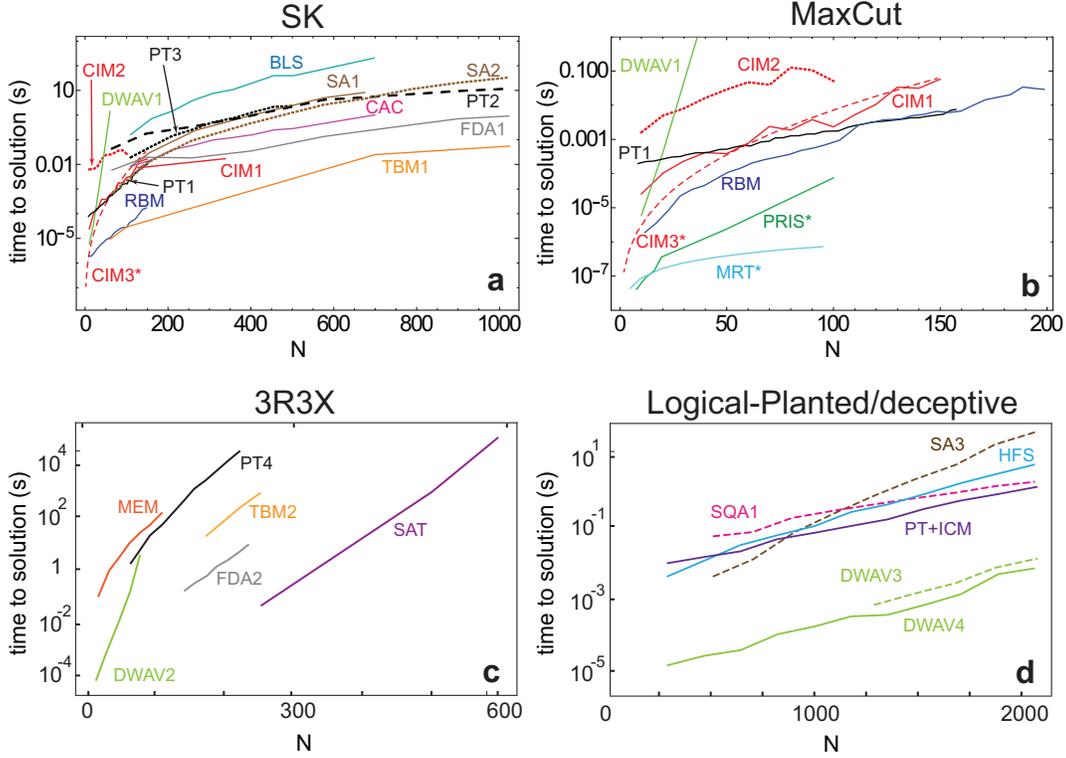} 
\end{center}
\caption{ {\bf Time-to-solution (TTS) comparison of various Ising machines. }  The time to obtain a 99\% success probability of obtaining the ground state is shown for the (a) SK; (b) dense MaxCut problems; \change{(c) 3R3X problems (reproduced from Ref. \cite{kowalsky20213}); (d)  logical-planted (dashed lines, from Ref. \cite{albash2018demonstration}) and deceptive cluster loops (solid lines, from Ref. \cite{mandra2018deceptive}) instance classes.  For the SK and MaxCut cases, the  Ising model definitions are as in Fig. \ref{fig4}.  The 3R3X, logical planted, and crafted problem definitions can be found in Refs. \cite{kowalsky20213},\cite{albash2018demonstration},\cite{ mandra2018deceptive} respectively. }  The labels for each line and their references are given in Table \ref{table1}. \change{Error bars on original data where present have been omitted for clarity.} For both SK and MaxCut, CIM1, CIM2, CIM3, DWAV and RBM are benchmarked on the same problem instances.  We note that for each reference, the best time to solution quoted are taken for each $ N $.  For results showing multiple annealing times, we have taken results optimized over annealing times. \change{Data reported for theoretical predictions, rather than being directly measured from an hardware implementation, are labeled with *.}  
\label{fig5}} 
\end{figure}

In Fig. \ref{fig4}, we show the performance of various Ising machines, quantified by the success probability for random instances of Sherrington-Kirkpatrick (SK) problems and dense MaxCut problem instances.  
We note that while the same types of models are used for the comparison in Fig.  \ref{fig4} and  \ref{fig5},  the same problem instances were not necessarily used, as we have compiled results from different studies. While this does not make our comparisons perfect, we hope that this nevertheless gives a sense of state of the art of various approaches. In numerous works,  the general scaling behavior is observed to follow the relation
\begin{align}
    p_{\text{suc}} \propto e^{-b N} ,
\end{align}
where $ b $ is a fitting parameter. Keeping in mind the interpretational caveat mentioned above about how the success probability can for some Ising-machine approaches been made high at the expense of very long runtimes in a way that is ultimately not useful, Fig.~\ref{fig4} gives evidence suggestive that at current technological levels, \change{SA}-based approaches, such as the restricted Boltzmann machines (RBM), implemented on digital hardware give the best scaling with $ N $.  One of the reasons for the high success probability of RBM is the inherent parallelism of this architecture which allows parallel SA updating. 
\change{We do however note that this figure does not include results from state-of-the-art dynamical-systems algorithms such as the CIM with amplitude-heterogeneity correction \cite{leleu2019destabilization,leleu2020chaotic} and SB \cite{goto2021high}, and based on these algorithms' excellent performance on the G-set MaxCut instances, one may anticipate that they would be competitive with RBM-based solvers. }
It is notable that the quantum annealer has a particularly poor performance in comparison to other methods.  This can be understood\cite{hamerly2019experimental} as a consequence of the benchmarked D-Wave machine having qubit connectivity given by a low-degree (Chimera) graph that cannot natively implement either the dense MaxCut or SK models (see Table \ref{table1}).  An embedding procedure that requires $ \propto N^2$ physical qubits is used to realize the equivalent graph, and this puts the D-Wave annealer at a disadvantage in comparison to the other listed approaches featuring all-to-all spin connections.  It is for this reason that the success probability has a relation which more resembles $ p_{\text{suc}} \propto e^{-b N^2} $.  


\begin{table}
  \resizebox{\textwidth}{!}{\begin{tabular}{|l|l|l|l|l|l|l|l|}
    \hline
   \hline
 \multirow{2}{*}{Ising machine/Algorithm}  & \multirow{2}{*}{Acronym}  &\multirow{2}{*}{Operating principle} & \multirow{2}{*}{Hardware}  & Hardware  & \multirow{2}{*}{Parallelization}  & Benchmark & \multirow{2}{*}{Reference} \\
 &  &  &  &  connectivity &  & problem &  \\
    \hline
        \hline
 Breakout local search & BLS & Local search \&  simulated annealing algorithm &  CPU &   All-to-all &  No & SK & Fig. 3a \cite{leleu2020chaotic} \\
   \hline
     Chaotic amplitude control & CAC &   Dynamical chaotic algorithm &     FPGA & All-to-all & Yes &  SK &  Fig. 3a \cite{leleu2020chaotic} \\
   \hline
  Coherent Ising machine (NTT) & CIM1 & Dynamical oscillator & \change{Hybrid (Optical/FPGA)} & All-to-all & Yes & MaxCut, SK &  Fig. S6 \cite{hamerly2019experimental} \\
   \hline
    Coherent Ising machine (Stanford) & CIM2 &    Dynamical oscillator & \change{Hybrid (Optical/FPGA)} &  All-to-all & Yes & MaxCut, SK & Fig. S6 \cite{hamerly2019experimental} \\
       \hline
   Coherent Ising machine & CIM3 &  Dynamical oscillator algorithm &  \change{*Predicted} &  All-to-all & Yes &  MaxCut, SK & Fig. S10 \cite{hamerly2019experimental} \\  
   \hline
    D-Wave quantum annealer 2Q& DWAV1 &   Quantum annealer  & \change{Superconducting qubits} &  Chimera & Yes & MaxCut, SK &  Fig. 3b, 4c \cite{hamerly2019experimental} \\
 \hline
  \change{D-Wave quantum annealer Advantage1.1} & DWAV2 &   Quantum annealer  & \change{Superconducting qubits} &  Chimera & Yes & 3R3X & Fig. 2 \cite{kowalsky20213} \\
   \hline
  \change{D-Wave quantum annealer 2KQ} & DWAV3 &   Quantum annealer  & \change{Superconducting qubits} &  Chimera & Yes & LP & Fig. 2 \cite{albash2018demonstration} \\
  \hline
  \change{D-Wave quantum annealer 2KQ} & DWAV4 &   Quantum annealer  & \change{Superconducting qubits} &  Chimera & Yes & Deceptive & Fig. 1 \cite{mandra2018deceptive} \\
    \hline
    Fujitsu digital annealer &  FDA1 & Simulated annealing algorithm &  ASIC &  All-to-all & Yes & SK & Fig. 7a \cite{aramon2019physics} \\
   \hline
\change{Fujitsu digital annealer} &  FDA2 & Simulated annealing algorithm &  ASIC &  All-to-all & Yes & 3R3X & Fig. 2 \cite{kowalsky20213}  \\
   \hline
\change{Hamze–de-Freitas–Selb} &  HFS &Tree sampling &  CPU &  All-to-all & No & Deceptive & Fig. 1\cite{mandra2018deceptive}  \\
   \hline
\change{Memcomputing} & MEM  & Dynamical logic gate algorithm & \change{CPU} & All-to-all &  Yes &  3R3X & Fig. 2 \cite{kowalsky20213} \\
   \hline 
   Memristor annealing  & MRT &   Simulated annealing algorithm & \change{*Predicted} &  All-to-all  & Yes & MaxCut &  Fig. 6a, 6b  \cite{cai2020power} \\
   \hline
       Photonic recurrent Ising sampler & PRIS &     Oscillator based annealer & \change{*Predicted} &  All-to-all   &  Yes & MaxCut & Fig. 2b \cite{roques2020heuristic} \\
          \hline
    Parallel tempering  & PT1 &  Simulated annealing algorithm &  CPU &  All-to-all & No & MaxCut, SK & Fig. S12 \cite{hamerly2019experimental} \\
   \hline
     Parallel tempering & PT2 &   Simulated annealing algorithm &  CPU & All-to-all & No  &  SK & Fig. 7a \cite{aramon2019physics} \\
   \hline
     Parallel tempering & PT3 &   Simulated annealing algorithm &  CPU &  All-to-all & No & SK & Fig. 3a \cite{leleu2020chaotic} \\
 \hline
      \change{Parallel tempering} & PT4 &   Simulated annealing algorithm & CPU &  All-to-all & No & 3R3X & Fig. 2 \cite{kowalsky20213}\\
  \hline
    \change{Isoenergetic cluster moves+ parallel tempering} & PT+ICM & Monte Carlo algorithm & Digital- CPU & All-to-all & Yes &  Deceptive & Fig. 1 \cite{mandra2018deceptive} \\
 \hline
     Restricted Boltzmann machine & RBM &  Simulated annealing algorithm &  FPGA & All-to-all & Yes &  MaxCut,SK & Fig. 3, 4 \cite{patel2020ising} \\
        \hline
       Simulated annealing & SA1 &    Simulated annealing algorithm &  CPU & All-to-all & Yes &   SK & Fig. 3a \cite{leleu2020chaotic} \\
   \hline
     Simulated annealing & SA2 &     Simulated annealing algorithm &  CPU & All-to-all &  No  &  SK & Fig. 7a \cite{aramon2019physics} \\
   \hline
        Simulated annealing     & SA3 &  Simulated annealing algorithm &  GPU & All-to-all &  Yes  &  LP & Fig. 2  \cite{albash2018demonstration} \\
   \hline
SATonGPU    & SAT &  SAT algorithm & GPU & All-to-all &  Yes  &  3R3X & Fig. 2 \cite{kowalsky20213} \\
   \hline
      Simulated quantum annealing & SQA1 & Quantum Monte Carlo algorithm  &  GPU & All-to-all & Yes   &  LP &  Fig. 2  \cite{albash2018demonstration} \\
    \hline
  Toshiba bifurcation machine & TBM1 &  \change{Discrete simulated bifurcation algorithm} &  FPGA & All-to-all &  Yes &  SK & Fig. 3c \cite{goto2021high} \\
  \hline 
  \change{Toshiba bifurcation machine} & TBM2  & \change{Discrete simulated bifurcation algorithm}   &  GPU & All-to-all &  Yes &  3R3X & Fig. 2 \cite{kowalsky20213} \\
   \hline
  \end{tabular}}
\caption{{\bf Types of Ising machine examined in Figs. \ref{fig4} and \ref{fig5}.  }  For each type of Ising machine, the operating principle, hardware type, hardware connectivity, and the parallelization are shown.  \change{We note that if the results of Fig. \ref{fig4} and \ref{fig5} are for theoretical predictions, rather than being directly measured from an hardware implementation, the hardware type is quoted as being ``*Predicted''. }   \change{In the Parallelization column we indicate approaches where simultaneous updates of Ising spins are performed. The Reference for publication source of the plotted data is given.  }  
\label{table1}
}
\end{table}

Figure \ref{fig5}a and \ref{fig5}b shows the time-to-solution metric for the MaxCut and SK models.  The best performing methods for the SK model employ classical digital hardware, where RBM and TBM show the lowest time to solution. \change{For MaxCut problem, the lowest time-to-solution achieved for a physically implemented machine is RBM.  We note that the MRT, PRIS, and CIM3 curves involve theoretical prediction of the time-to-solution, rather than a direct measurement of the time. }  The scaling relation for most of the curves  follows the phenomenological relation
\begin{align}
    T_{\text{sol}}  \propto e^{c\sqrt{N}} .
    \label{squareroottts}
\end{align}
\change{For the cases with a relatively small range of available data the square root behavior may not yet be visible.}
\change{The D-wave results are better approximated by an exponential relation $ T_{\text{sol}}  \propto e^{cN} $ which requires $ \propto N^2$ physical qubits due to the limited chimera connectivities of the qubits.} One should note that this is a hardware implementation limitation that gives a different scaling and not the computational mechanism itself, and may be improved in the  future  \cite{mukai2019superconducting,onodera2020quantum,lechner2015quantum,puri2017quantum}. For problem instances with sparse connectivity D-wave was observed to have a more favorable scaling \cite{hamerly2019experimental}.  These results show that the connectivity is an important factor that determines the performance of an Ising machine --- in Table. \ref{table1}, the hardware connectivity of the different Ising machines are given.

\change{In Fig. \ref{fig5}c, we show results from Ref. \cite{kowalsky20213}, which compare various Ising machines for 3R3X problems, which have a golf-course energy landscape structure with known exact solutions. The best performing approach in this case is the SATonGPU approach, which is a highly parallelized version of a SAT algorithm implemented on a GPU.  The Fujitsu Digital annealer and Toshiba bifurcation machine achieves almost similar scaling although has a larger prefactor than the SATonGPU approach.  The memcomputing results are based on classical simulation of a proposed system, hence with dedicated hardware to implement it, some performance improvement might result \cite{kowalsky20213}. Although there are fewer studies performed for this problem class, these results again suggest that the best performing solvers today are based on digital computing hardware. }

To show the potential of quantum approaches, we also discuss additional problem classes where it is expected that QA has advantages over a class of classical methods \cite{albash2018demonstration,denchev2016computational,boixo2016computational} despite the mentioned limitations of D-wave.  In Fig. \ref{fig5}d,  the optimum time-to-solution for the class of logical-planted (LP) problems that are constructed such that they promote the presence of tunneling barriers is compared (dashed lines). For these problems it is expected that barriers can be traversed more effectively by quantum, rather than thermal fluctuations.  Here D-Wave and SQA shows a scaling advantage over SA \cite{albash2018demonstration}. 
 \change{The superior performance of SQA  implies that tunneling through barriers may not be considered the exclusive advantage of quantum hardware. However, one should note that that SQA can not be applied for non-stoquastic Hamiltonians where there is a sign problem, and as such the power of QA for non-stoquastic Hamiltonians require further exploration. Non-stoquastic Hamiltonians are important from a computational complexity perspective as adiabatic quantum computation with non-stoquastic Hamiltonians is equivalent to the circuit model of quantum computing \cite{aharonov2008adiabatic}.  Therefore, they can simulate other universal models with at most a polynomial resource overhead.  The fact that D-Wave outperforms SA confirms the presence and advantage of quantumness but the superior performance of SQA suggests that
current QA hardware is still dominated by classical dynamics and needs to be improved.} \change{We note that for the LP problem class, there are classical algorithms that outperform or have comparable performance with D-Wave \cite{albash2018demonstration}. In Fig. \ref{fig5}d (solid lines), we show results from a separate study \cite{mandra2018deceptive} that compares D-wave with classical heuristic algorithms for another class of specially crafted problem, called the deceptive cluster loop problem. For this problem class, D-wave outperforms the best known heuristics algorithms such as parallel tempering Monte Carlo with isoenergetic cluster moves (PT+ICM) and Hamze–de Freitas–Selby (HFS) with approximately two orders of magnitude shorter TTS. However, no scaling improvement is evident. } 

\change{Numerous other benchmarking studies of Ising machines have been performed in the literature (see for e.g. Refs. \citen{hamerly2019experimental,oshiyama2021benchmark,cai2020power,king2018emulating,aramon2019physics,leleu2020chaotic,goto2019combinatorial,goto2021high,patel2020ising,harrigan2021quantum}). For example, in Ref. \cite{oshiyama2021benchmark}, the performance of the D-Wave hybrid solver, TBM, FDA, and SA was benchmarked for three different classes of problem instances including SK. The results highlight  the fact that the performance of machines is problem dependent. In particular for SK model, TBM showed the best performance. In Ref. \cite{harrigan2021quantum}, the performance of QAOA was benchmarked on SK and MaxCut problems for problems up to 23 qubits. }


\section*{Discussion and Outlook}

We have surveyed various approaches to Ising machines and their computational performance.  We identify three dominant operating-principle strategies that Ising machines employ: classical annealing, quantum annealing, and dynamical system evolution.  All three approaches are all in principle compatible with the natural-computing approach, where information is encoded in an analog and parallelized way such that the underlying physics drives the system towards the ground state.

In the dynamical-system approach, the evolution drives the state of the system far from equilibrium, in contrast to annealing approaches where the aim is to avoid occupying high-energy states.  In classical and quantum annealing, adding thermal or quantum noise is an important component of the procedure to escape local minima in the energy landscape.  On the other hand, in dynamical approaches, the system is designed to be attracted to particular configurations that correspond to low-energy configurations of a given Ising instance. Comparing the performance of Ising machines, interestingly, most approaches tend to have similar scalings in terms of the error probability and the time-to-solution metrics as a function of the number of spins, despite extremely different approaches and technologies used to realize them.  The complexity of all approaches scale exponentially with the system size, with the difference being the power within the exponent and the prefactors. This is expected given the NP-complete complexity of the Ising problem---the battle between competing approaches is with respect to the exponents that are achievable, where a small difference in the exponent makes a large difference in time-to-solution for large system sizes. 

While Fig. \ref{fig4} and \ref{fig5} suggest that classical digital methods are still the best-performing approaches at the time of writing; analog and quantum computing technologies are rapidly developing and the landscape may completely change in a short amount of time.  Some of the best-performing approaches are based on classical digital technology which have had the benefit of decades of development, and in many cases can be highly parallelized.  
In comparison, QA approaches have only been recently developed to a scale where it can be tested, either theoretically or experimentally, and often have hardware limitations such as limited connectivity and the presence of decoherence. The specific form of the Ising instance being solved (e.g., the structure of the $J$ matrix) can affect the performance dramatically.  It may be that in the future, much like various numerical algorithms are chosen based on the compatibility of a particular problem, that different Ising machines will be utilized according to their suitability for the given problem. For example, while problem instances with small spectral gaps are known to be hard for QA, they may not limit the performance of QAOA \cite{zhou2020quantum}. 
One step further would be to explore hybrid quantum-classical and digital-analog algorithms to gain the complementary advantage of each \cite{chancellor2017modernizing}. 

One point of active debate has been the role that quantum mechanics plays in coherent Ising machines and in D-Wave's quantum annealers. In the context of coherent Ising machines, there exist models of their operation that treat CIMs quantum mechanically\cite{wang2013coherent,yamamoto2017coherent}, including, for example, a description of the initial state as being in a coherent superposition of all logical states\cite{yamamoto2017coherent}. However, experimental realizations of CIMs thus far\cite{marandi2014network,inagaki2016large,mcmahon2016fully,inagaki2016coherent,honjo2021spin} have been in regimes of high photon loss, where purely classical models can accurately describe the pertinent dynamics of the systems. The clearest indication of this is that similar performance to demonstrated coherent Ising machines  may be achieved by simulating the mean-field dynamics \cite{bilbro1989optimization,king2018emulating,tiunov2019annealing,hamerly2019experimental}. With sufficiently high nonlinearity to loss in their constituent OPOs, CIMs can be firmly in the quantum regime\cite{onodera2018nonlinear} and have a strong connection with quantum annealers\cite{goto2016bifurcation}. Exploring how to construct experimental CIMs where quantum effects play a crucial role, and designing them so that quantum effects improve the performance of the machine, are two topics of active investigation \cite{yamamoto2020coherent}.  

For QA, the computational advantage of incoherent tunneling over certain classical methods (typically \change{SA}\cite{kirkpatrick1983optimization} and the Hamze--de~Freitas--Selby algorithm\cite{hamze2012fields,selby2014efficient}) for certain classes of problems has been shown \cite{RevModPhys.90.015002,albash2018demonstration,PhysRevX.6.031010,denchev2016computational,boixo2016computational,mandra2018deceptive}. However, for any real-world problem of interest, no evidence of an unqualified quantum speedup (as defined in Ref.~\cite{job2018test}) has been found. Perhaps the most compelling results with the D-Wave QA so far are for a specially crafted problem class, \textit{deceptive cluster loops}, for which the QA was found to outperform in terms of time-to-solution for all classical heuristics that were tested, including parallel tempering \cite{zhu2015efficient}. The speedup was of an approximately constant-factor nature, with no strong evidence of a scaling advantage \cite{mandra2018deceptive}. \change{Another disadvantage of quantum annealing is that they cannot sample uniformly all low-lying states in contrast to other heuristic algorithms such as SA-based algorithms \cite{mandra2017exponentially,zhu2019fair}.  In SA, after many repetitions and starting from different initial states one can record all the configurations that minimize the problem Hamiltonian.  For an optimization machine, such an ability to sample fairly is beneficial as having different solutions for a problem is often useful. } Furthermore, it is not yet well-understood what the role of entanglement in QA is and whether it contributes to a quantum speedup \cite{albash2018adiabatic}.  While various aspects of quantumness, including entanglement, might or might not aid the performance of CIMs or QAs, this uncertainty has inspired the proposal of interesting quantum-inspired classical algorithms related to CIMs and QAs, which is a fruitful development in its own right.

Looking to the future,  there is much room for development for Ising machines.  In the same way that the scaling of the time-to-solution and other metrics for classical algorithms are known to several decimal places, the scaling of Ising machines require more precise quantification such that competing methods can be compared.  Figs. \ref{fig4} and \ref{fig5} are extremely preliminary in this regard. With improved quantification and investigation for different classes of problems, a better understanding of the suitability of various approaches for a particular problem can be known in advance. 
Another interesting direction  is to compare the performance of the different Ising machines for finding an approximate solution with different levels of accuracy, since for many applications finding a high-quality solution, rather than the exact solution, is sufficient.  
Several forms of Ising machine whose operation either relies on or can be enhanced by quantum-mechanical mechanisms have been proposed and demonstrated. However, constructing large-scale quantum machines with high connectivity and low decoherence remains an outstanding challenge for the field of quantum information processing in general, and further progress in this direction is needed for experimental exploration of the benefits quantum-mechanical methods may bring to solving Ising problems. \change{This is in contrast with classical approaches, especially digital ones, which often have little difficulty in supporting full connectivity.}
Given the demand for faster methods of solving optimization problems in society, and the maturity of conventional algorithms and processors, it seems likely that the development of specialized Ising machines will continue well into the future, featuring an exciting interplay between hardware engineering, computer science, statistical physics, and quantum mechanics.


\section*{Acknowledgements}
We thank Scott Aaronson, Tameem Albash, Helmut Katzgraber, Timoth{\'e}e Leleu, Sam King, Marek Narozniak, Srikrishna Vadlamani and Thomas van Vaerenbergh for helpful discussions and comments on the manuscript.  T.B. is supported by the National Natural Science Foundation of China (62071301); NYU-ECNU Institute of Physics at NYU Shanghai; the Joint Physics Research Institute Challenge Grant; the Science and Technology Commission of Shanghai Municipality (19XD1423000,22ZR1444600); the NYU Shanghai Boost Fund; the China Foreign Experts Program (G2021013002L); the NYU Shanghai Major-Grants Seed Fund. P.L.M. thanks all his collaborators on the topic of Ising machines---especially Surya Ganguli, Ryan Hamerly, Timothée Leleu, Hideo Mabuchi, Alireza Marandi, Edwin Ng, Tatsuhiro Onodera, and Yoshihisa Yamamoto---for enlightening discussions that have shaped his understanding over the years. P.L.M. acknowledges funding from NSF award CCF-1918549, and NTT Research for their financial and technical support. P.L.M. also acknowledges membership in the CIFAR Quantum Information Science Program as an Azrieli Global Scholar.

\section*{Author contributions}
All authors contributed in compiling the results and preparing the manuscript.  

\section*{Competing interests}
P.L.M. declares an interest in QC Ware Corp., a company producing software for quantum computers, to which he is an advisor.  T.B. and N.M. declare no competing interests.

\section*{Author's note}
This is a preprint of a review paper that is scheduled to appear in Nature Reviews Physics. This version does not include changes made during the editing and production process at the journal.

\end{document}